\documentclass[10pt,journal,comsoc]{IEEEtran}

\usepackage{cite}
\usepackage{graphicx}
\usepackage{amsmath,amssymb}
\usepackage{url}
\usepackage[table]{xcolor}
\usepackage{booktabs}
\usepackage{tabularx}
\usepackage{array}
\usepackage{ragged2e}
\usepackage{subcaption}
\usepackage{tikz}
\usetikzlibrary{arrows.meta,positioning,calc}
\usepackage{xcolor}
\usepackage{orcidlink}
\usepackage{hyperref}

\begin{document}

\title{Adaptive UAV Communications for URLLC: From Preplanned Designs to Real-Time Intelligence}

\author{Asim Ihsan \,\orcidlink{0000-0001-7491-7178},
Muhammad Asif \,\orcidlink{0000-0002-9699-1675},
Ali Arshad Nasir\,\orcidlink{0000-0001-5012-1562}, Senior Member, IEEE, Wali Ullah Khan\,\orcidlink{0000-0003-1485-5141}, and
Khaled M. Rabie\,\orcidlink{0000-0002-9784-3703}, Senior Member, IEEE
\thanks{Asim Ihsan, Ali Arshad Nasir, and Khaled M. Rabie are with the Interdisciplinary Research Center for Communication Systems and Sensing (IRC-CSS), King Fahd University of Petroleum and Minerals (KFUPM), Saudi Arabia. Ali Arshad Nasir is also with the Department of Electrical Engineering, and Khaled M. Rabie is also with the Department of Computer Engineering.}
\thanks{Muhammad Asif is with the School of Electrical Engineering, Tongling University, China.}
\thanks{Wali Ullah Khan is with the Interdisciplinary Centre for Security, Reliability, and Trust, University of Luxembourg, Luxembourg.}}
%\thanks{Author One and Author Two are with ...}
%}

\maketitle

\begin{abstract}
Unmanned aerial vehicles (UAVs) are emerging as a key enabler of next-generation wireless networks, particularly for applications that require ultra-reliable and low-latency communication (URLLC), such as emergency response, industrial automation, and autonomous systems. In these scenarios, maintaining reliable connectivity under strict transmission time constraints is challenging due to dynamic environments, mobility, and limited onboard energy. In particular, communication performance and energy are closely coupled with UAV movement, making trajectory design a critical component of system operation. Most existing approaches rely on offline joint communication and trajectory optimization, where the UAV trajectory and communication parameters are optimized prior to execution based on assumed system information. Although effective under ideal assumptions, such designs cannot adapt to real-time variations in user demand, channel conditions, or environmental disturbances, which are particularly critical in URLLC settings. To address these challenges, this article investigates model predictive control (MPC) as an adaptive framework for UAV-enabled communications. Using a receding-horizon strategy, MPC enables the UAV to continuously update its trajectory based on real-time information, improving reliability and robustness in dynamic environments. Representative application scenarios are discussed to highlight the role of MPC in UAV-enabled URLLC systems. Furthermore, a case study is  presented to illustrate key design trade-offs and performance insights under finite blocklength-based URLLC transmission, followed by a discussion on open challenges and future research directions for practical and scalable MPC-enabled UAV communication systems.
\end{abstract}

\section{Introduction}
Unmanned aerial vehicles (UAVs) have emerged as a transformative technology in modern wireless networks, offering flexible deployment and adaptive coverage in dynamic environments. Their applications range from emergency response and environmental monitoring to next-generation 5G and beyond communication systems. However, unlike conventional terrestrial infrastructure, UAV-enabled communication systems introduce a unique interplay between mobility, communication performance, and energy consumption.

A fundamental aspect of UAV operation is trajectory planning, which involves determining the UAV’s motion from an initial point to the desired final destination. Generally, trajectory planning has been explored by navigation and control engineers to achieve goals related to path tracking accuracy, obstacle avoidance, and system stability. Model predictive control (MPC) is widely used in trajectory planning due to its predictive nature, where it can forecast the future state of the system and optimize control actions under constraints in real time \cite{9938397}. This predictive capability makes MPC particularly suitable for UAV systems operating in dynamic and uncertain environments.

At the same time, studies on UAV-assisted communications have shown that trajectory design plays a critical role in determining communication performance. UAVs can dynamically adjust their positions to improve channel quality, establish line-of-sight (LoS) links, and enhance connectivity for ground users. Recent findings, for example, have indicated that UAV trajectory optimization can significantly improve communication reliability and coverage by adapting to environmental conditions and user locations\cite{10839492}.
\begin{figure*}[!t]
    \centering
    \includegraphics[width=\textwidth]{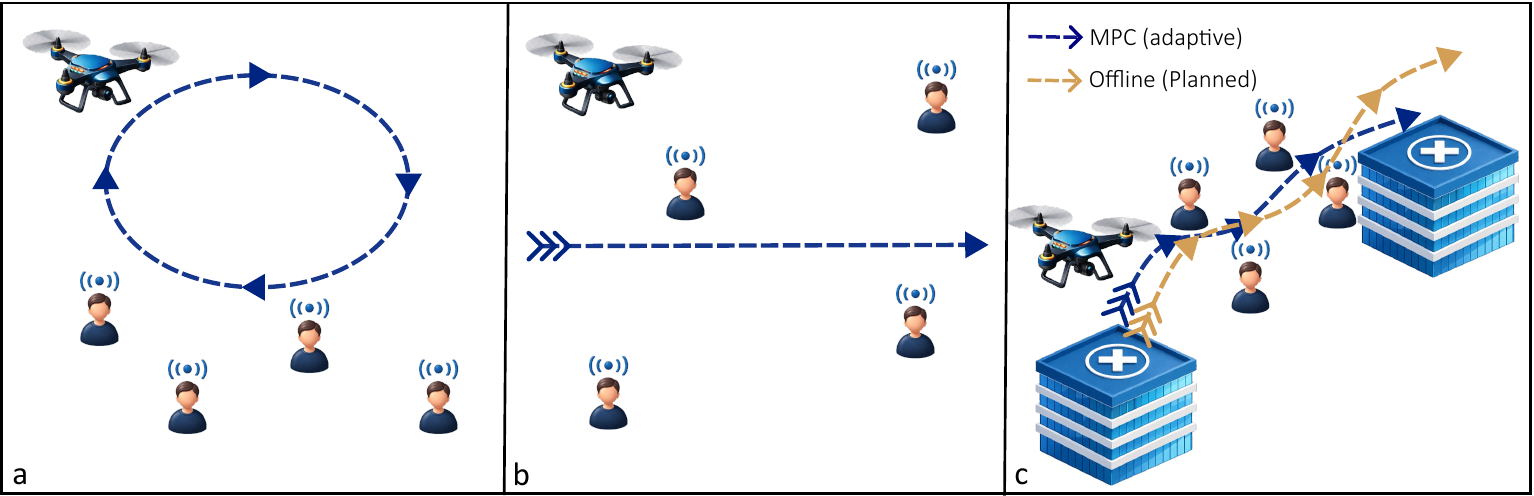}
    \caption{Evolution of UAV trajectory design paradigms. (a) Circular trajectory. (b) Straight-line trajectory. (c) Communication-aware trajectories under disturbances. In (c), the offline planned trajectory may deviate under disturbances, while the MPC-based adaptive trajectory continuously adjusts to reach the intended destination.}
    \label{fig:trajectory_evolution}
\end{figure*}
However, an important insight emerging from the literature is that trajectory design cannot be treated independently of energy consumption. UAVs are inherently energy-limited systems, where a significant portion of the energy is consumed by propulsion rather than communication. As highlighted in previous work, effective operation of UAV-assisted networks requires the joint consideration of transmission energy and propulsion energy, since both contribute to the overall cost of the system \cite{7888557}. This observation fundamentally changes how UAV trajectory optimization should be approached. In particular, the path taken by the UAV from its starting point to its destination is no longer simply the shortest or fastest route. Instead, it becomes a control variable that directly influences both communication efficiency and energy expenditure. Flying closer to users can improve channel conditions and reduce transmission power, but it often requires additional maneuvering, increasing propulsion energy. Conversely, energy-efficient flight paths may degrade communication quality, requiring higher transmit power to maintain performance.

This interplay becomes even more critical in communication scenarios requiring ultra-reliable and low-latency communication (URLLC). In such settings, communication must be maintained under strict reliability and timing requirements, where short packet transmission plays a dominant role. As a result, traditional communication models are no longer sufficient to capture the trade-off between reliability and transmission delay. In particular, under the finite blocklength (FBL) regime, achievable rates depend not only on signal-to-interference-plus-noise ratio (SINR) but also on blocklength and decoding error probability \cite{11214543}, making the joint communication and trajectory optimization significantly more challenging than conventional Shannon-based formulations.

MPC-based approaches have been applied to joint trajectory and communication design, typically with rate-oriented objectives \cite{11069265}. Extending these approaches to system-level energy efficiency that jointly accounts for communication and propulsion energy \cite{7888557}, as well as URLLC-driven design, introduces additional challenges. In particular, combining system-level energy efficiency and predictive trajectory control, while explicitly accounting for propulsion energy in trajectory dynamics, within a unified framework for URLLC application remains an open problem. To address this, this article presents a unified perspective on UAV-enabled communication systems, where trajectory design and system-level energy efficiency, accounting for both communication and propulsion power, are jointly considered under URLLC requirements. Using MPC, the proposed framework enables real-time adaptation of both UAV movement and transmission strategies, supporting energy-efficient operation under dynamic conditions while accounting for the trade-off between reliability and transmission delay.

The main contributions of this article are as follows. We present a system-level view of UAV-enabled communication and show how trajectory design, communication performance, and energy use are closely linked. We then review the shift from offline and geometric trajectory design to MPC-based adaptive control, and explain why fixed offline designs cannot handle dynamic conditions. We study energy-efficient communication under short-packet transmission in the FBL regime and highlight the trade-off between energy efficiency  and transmission delay in URLLC systems. We develop an MPC-based framework that jointly updates UAV trajectory and communication decisions in real time and apply it across system design, case study, and application scenarios. We also provide design insights on balancing system-level energy efficiency and trajectory progression. We finally discuss key challenges and future directions, with a focus on extending MPC for practical and scalable UAV-enabled URLLC systems.

\section{From Geometric Paths to Communication-Aware UAV Trajectories}
UAV trajectory design has evolved from simple geometric movement patterns to more intelligent, communication-aware strategies. As illustrated in Fig.~\ref{fig:trajectory_evolution}, this evolution spans from basic geometric paths to adaptive and predictive designs. In earlier research work, the UAVs had pre-defined trajectories, such as straight-line paths, for performing certain missions including data relay and delivery with little flexibility in adjusting the path according to the requirement of communication. Nevertheless, as the wireless system requirement increases, it has been realized that the trajectory of the UAV can be exploited to improve its communication performance. Therefore, modern approaches incorporate trajectory design with their communication objective, enabling UAVs to dynamically adjust their location for the improvement of communication reliability  through interference mitigation in complex and changing environments. 
\subsection{Straight-Line Trajectory: A Simplified Mobility Assumption}
Many prior studies simplify UAV mobility by assuming straight-line trajectories, in which the UAV moves along a predetermined path while delivering communication services. 
This simplification draws on real-world UAV operations such as infrastructure inspection and border surveillance, where the UAV follows a fixed route \cite{9122470}. The flight path of UAVs in these applications inherently exhibits a straight route, and communication takes place along that route. Additionally, modelling the trajectory as a straight line eases the analysis and serves as a benchmark in many studies to quantify the impact of UAV's mobility on communication performance. Therefore, a straight line trajectory is primarily applicable to theoretical investigations or application-driven studies, rather than general system design. 
\subsection{Circular Trajectory: Persistent Coverage and Service Continuity}
Circular trajectories are widely investigated in the literature because of their persistent coverage over a target area. The periodic nature of circular motion can facilitate more balanced service over time in multi-user settings, especially under well-designed scheduling strategies. Therefore, this type of UAV trajectory has been explicitly examined in the context of energy-efficient UAV communications. They optimized the energy efficiency of the UAV-enabled communication system as a function of  parameters such as flight radius and velocity of the UAV \cite{7888557}.
 Nevertheless, the inflexibility of circular motion makes it unsuitable for adapting to spatially diverse or uneven user distributions.
\subsection{Communication-Aware Trajectories: Mobility Driven by System Objectives}
Modern UAV communication frameworks increasingly incorporate communication-aware trajectory design, where UAV mobility is explicitly coupled with system performance metrics \cite{11390050}. In this paradigm, the UAV adapts to improve link quality, manage interference, and improve overall network efficiency. This concept has been extensively explored through joint optimization frameworks that integrate trajectory planning with communication strategies such as scheduling, power control, and beamforming. These studies consistently demonstrate that substantial performance gains can be achieved by exploring the UAV’s spatial degrees of freedom in a coordinated and communication-driven manner. Despite their performance gains, most communication-aware trajectory designs rely on offline optimization with prior knowledge of system parameters. In practice, UAV operations are affected by environmental disturbances such as wind and turbulence, as well as navigation and control inaccuracies \cite{9678336}. These disturbances can deviate the UAV from its planned trajectory, resulting in performance degradation, as illustrated in Fig.~\ref{fig:trajectory_evolution}(c). Therefore, offline designs are unreliable and insufficient for real-world deployment.
\subsection{Adaptive and Predictive Trajectories: Toward Real-Time UAV Operation}
Traditional UAV trajectory designs are mainly based on offline optimization. Recently, more attention has been given to adaptive approaches that update the UAV trajectory during operation \cite{11069265}. These methods allow the UAV to respond to changing network and environmental conditions in real time. MPC is one such approach that uses short-term prediction to optimize control actions over a finite horizon. This enables the UAV to adjust its movement while maintaining communication performance.
%\definecolor{headerblue}{RGB}{33,96,154}
%\definecolor{rowgray}{RGB}{245,245,245}

%\begin{table*}[t]
%\centering
%\caption{Trajectory Design Paradigms in UAV-Enabled Communications}
%\label{tab:trajectory_types}
%\renewcommand{\arraystretch}{1.25}
%\setlength{\tabcolsep}{5pt}
%\small
%\rowcolors{2}{rowgray}{white}
%\begin{tabularx}{\textwidth}{
%>{\RaggedRight\arraybackslash}p{2.5cm}
%>{\RaggedRight\arraybackslash}X
%>{\RaggedRight\arraybackslash}X
%>{\RaggedRight\arraybackslash}X
%>{\RaggedRight\arraybackslash}p{2cm}}
%\rowcolor{headerblue}
%\color{white}\textbf{Trajectory Type} &
%\color{white}\textbf{Main Strength} &
%\color{white}\textbf{Limitation} &
%\color{white}\textbf{Reference} \\
%\toprule
%Straight-Line &
%Baseline mobility model for communication analysis &
%Simple, interpretable, analytically tractable &
%Limited adaptability; suboptimal for multi-user scenarios &
%\cite{9122470} \\

%Circular &
%Persistent coverage over a service region &
%Stable link distances; energy-aware operation &
%Rigid geometry; limited adaptability to user distribution &
%\cite{7888557} \\

%Communication-Aware &
%Joint trajectory and communication optimization &
%Enhances SINR, throughput, and interference management &
%Higher computational complexity &
%\cite{11390050} \\

%Adaptive and Predictive &
%Real-time trajectory refinement based on system conditions &
%Robust to dynamics; supports time-varying environments &
%Requires real-time computation and system feedback &
%\cite{11069265} \\

%\bottomrule
%\end{tabularx}
%\end{table*}
\section{From Offline to Online Predictive UAV Trajectory Design}
\subsection{Offline Trajectory Design: Insightful but predefined and Static}
%%%%%%%%%%%%%%%%%%%%%%%%%%%%%
%========================================================
% Common color palette
%========================================================
\definecolor{flowbase}{RGB}{250,246,238}
\definecolor{flowaccent}{RGB}{220,232,246}
\definecolor{flowaccentdark}{RGB}{40,85,130}
\definecolor{flowborder}{RGB}{60,60,60}
\definecolor{flowtext}{RGB}{0,0,0}
\definecolor{feedbackfill}{RGB}{230,238,248}

%========================================================
% Figure 1: Offline UAV mission planning and execution framework
%========================================================
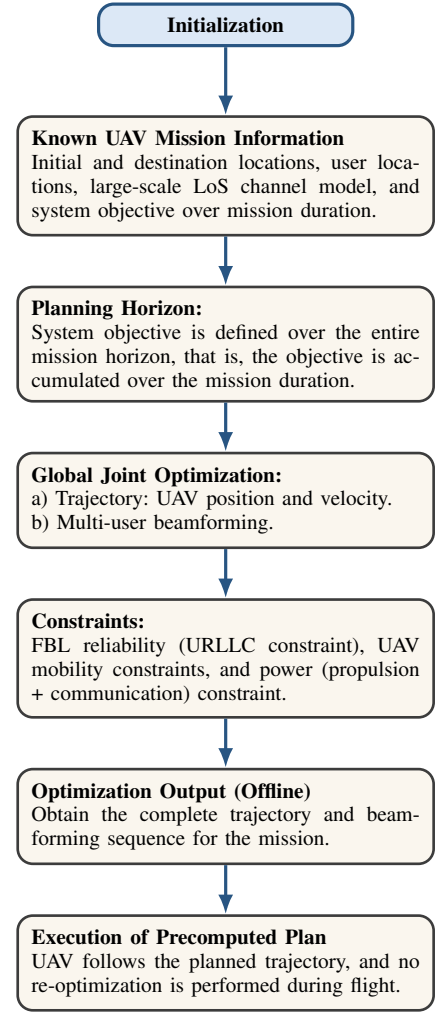
\begin{figure}[!t]
\centering
\footnotesize
\begin{tikzpicture}[
    node distance=7mm,
    >={Latex[length=2.6mm,width=2.0mm]},
    box/.style={
        rectangle,
        rounded corners=2mm,
        draw=flowborder,
        line width=1.0pt,
        fill=flowbase,
        text=flowtext,
        align=justify,
        text width=0.58\columnwidth,
        inner sep=5.2pt
    },
    topbox/.style={
        rectangle,
        rounded corners=2mm,
        draw=flowaccentdark,
        line width=1.2pt,
        fill=flowaccent,
        text=flowtext,
        align=center,
        text width=0.34\columnwidth,
        inner sep=5pt,
        font=\bfseries
    },
    arrow/.style={
        draw=flowaccentdark,
        line width=1.2pt,
        -{Latex[length=2.6mm,width=2.0mm]}
    }
]

%--------------------------------------------------------
% Nodes
%--------------------------------------------------------
\node[topbox] (n1) {Initialization};

\node[box, below=9mm of n1] (n2) {%
\textbf{Known UAV Mission Information}\\
Initial and destination locations, user locations, large-scale LoS channel model, and system objective over mission duration.
};

\node[box, below=6.5mm of n2] (n3) {%
\textbf{Planning Horizon:}\\
System objective is defined over the entire mission horizon, that is, the objective is accumulated over the mission duration.
};

\node[box, below=6.5mm of n3] (n4) {%
\textbf{Global Joint Optimization:}\\
a) Trajectory: UAV position and velocity.\\
b) Multi-user beamforming.
};

\node[box, below=6.5mm of n4] (n5) {%
\textbf{Constraints:}\\
FBL reliability (URLLC constraint), UAV mobility constraints, and power (propulsion + communication) constraint.
};

\node[box, below=6.5mm of n5] (n6) {%
\textbf{Optimization Output (Offline)}\\
Obtain the complete trajectory and beamforming sequence for the mission.
};

\node[box, below=6.5mm of n6] (n7) {%
\textbf{Execution of Precomputed Plan}\\
UAV follows the planned trajectory, and no re-optimization is performed during flight.
};

%--------------------------------------------------------
% Arrows
%--------------------------------------------------------
\draw[arrow] (n1) -- (n2);
\draw[arrow] (n2) -- (n3);
\draw[arrow] (n3) -- (n4);
\draw[arrow] (n4) -- (n5);
\draw[arrow] (n5) -- (n6);
\draw[arrow] (n6) -- (n7);

\end{tikzpicture}
\caption{Offline UAV mission planning and execution framework.}
\label{fig:offline_flowchart}
\end{figure}

Most of the existing studies on UAV-enabled communication systems rely on offline trajectory design, where the UAV path is optimized before deployment based on the available system information such as user locations and channel conditions. These approaches usually rely on statistical channel and mobility models to capture the variation in the wireless channel and the user mobility. However, offline methods are based on statistical system behavior rather than real-time observations. Many studies assume the user locations or mobility patterns are known or can be estimated in advance \cite{mahmood2025uav}. Consequently, the UAV trajectory is fixed at execution time and cannot adapt to instantaneous deviations caused by wind disturbances, localization errors, or unexpected changes in the environment.

\subsubsection{How It Works}
Offline trajectory design plans the UAV movement before the mission begins. As shown in Fig.~\ref{fig:offline_flowchart}, the process starts by collecting available mission information, such as the initial and destination locations, user distribution, and channel distribution knowledge. Based on this information, an optimization problem is formulated over the entire mission horizon, where the objective is evaluated across all time steps. This allows the UAV trajectory and communication variables to be jointly optimized in a global manner. The result of this optimization is a complete trajectory and transmission strategy for the entire mission. During operation, the UAV follows this precomputed plan without updating it, even if environmental conditions change.

\subsection{MPC-Based Trajectory Design: Adaptive and Online Operation}
To overcome the limitations of precomputed offline trajectories, recent research has increasingly focused on adaptive trajectory design approaches, such as MPC \cite{11069265}. Unlike offline methods, which rely on preplanned paths, MPC allows the UAV to make decisions based on its current situation, taking into account real-time variations in user demand, channel conditions, and environmental disturbances. This capability is of great importance for UAV to maintain reliable connectivity in a  dynamic environment while progressing towards its mission objective.  
\subsubsection{How It Works}
\begin{figure}[!t]
\centering
\footnotesize
\begin{tikzpicture}[
    node distance=7mm,
    >={Latex[length=2.6mm,width=2.0mm]},
    box/.style={
        rectangle,
        rounded corners=2mm,
        draw=flowborder,
        line width=1.0pt,
        fill=flowbase,
        text=flowtext,
        align=justify,
        text width=0.58\columnwidth,
        inner sep=5.2pt
    },
    topbox/.style={
        rectangle,
        rounded corners=2mm,
        draw=flowaccentdark,
        line width=1.2pt,
        fill=flowaccent,
        text=flowtext,
        align=center,
        text width=0.34\columnwidth,
        inner sep=5pt,
        font=\bfseries
    },
    sidebox/.style={
        rectangle,
        rounded corners=2mm,
        draw=flowaccentdark,
        line width=1.1pt,
        fill=feedbackfill,
        text=flowtext,
        align=center,
        text width=0.24\columnwidth,
        inner sep=5pt,
        font=\bfseries
    },
    arrow/.style={
        draw=flowaccentdark,
        line width=1.2pt,
        -{Latex[length=2.6mm,width=2.0mm]}
    },
    feedbackarrow/.style={
        draw=flowaccentdark,
        line width=1.2pt,
        dashed,
        -{Latex[length=2.6mm,width=2.0mm]}
    }
]

%--------------------------------------------------------
% Main vertical flow
%--------------------------------------------------------
\node[topbox] (n1) {Initialization};

\node[box, below=9mm of n1] (n2) {%
\textbf{Current System State}\\
UAV position and velocity, large-scale LoS channel model, FBL-based user data rates, and energy efficiency (performance metric).
};

\node[box, below=6.5mm of n2] (n3) {%
\textbf{Prediction:}\\
System objective is over prediction horizon at each UAV time step, that is, the summation of objective is over prediction horizon.
};

\node[box, below=6.5mm of n3] (n4) {%
\textbf{Joint Optimization:}\\
a) Trajectory: UAV position and velocity.\\
b) Multi-user beamforming.
};

\node[box, below=6.5mm of n4] (n5) {%
\textbf{Constraints:}\\
FBL reliability (URLLC constraint), UAV mobility constraints, and power (propulsion + communication) constraint.
};

\node[box, below=6.5mm of n5] (n6) {%
\textbf{Control Action (Output)}\\
Apply only first UAV step from prediction horizon and Discard rest.
};

\node[box, below=6.5mm of n6] (n7) {%
\textbf{System state Update}\\
System state update based on control action + disturbance.
};

%--------------------------------------------------------
% Feedback box
%--------------------------------------------------------
\node[sidebox, right=31mm of $(n5)!0.5!(n4)$] (fb) {Updated State\\Feedback};

%--------------------------------------------------------
% Vertical arrows in main flow
%--------------------------------------------------------
\draw[arrow] (n1) -- (n2);
\draw[arrow] (n2) -- (n3);
\draw[arrow] (n3) -- (n4);
\draw[arrow] (n4) -- (n5);
\draw[arrow] (n5) -- (n6);
\draw[arrow] (n6) -- (n7);

%--------------------------------------------------------
% Feedback loop
%--------------------------------------------------------
\draw[feedbackarrow] (n7.east) -- ++(14mm,0) -| (fb.south);
\draw[feedbackarrow] (fb.north) |- ([xshift=1.5mm]n2.east);

\end{tikzpicture}
\caption{MPC-based UAV trajectory and beamforming optimization framework.}
\label{fig:mpc_flowchart}
\end{figure}
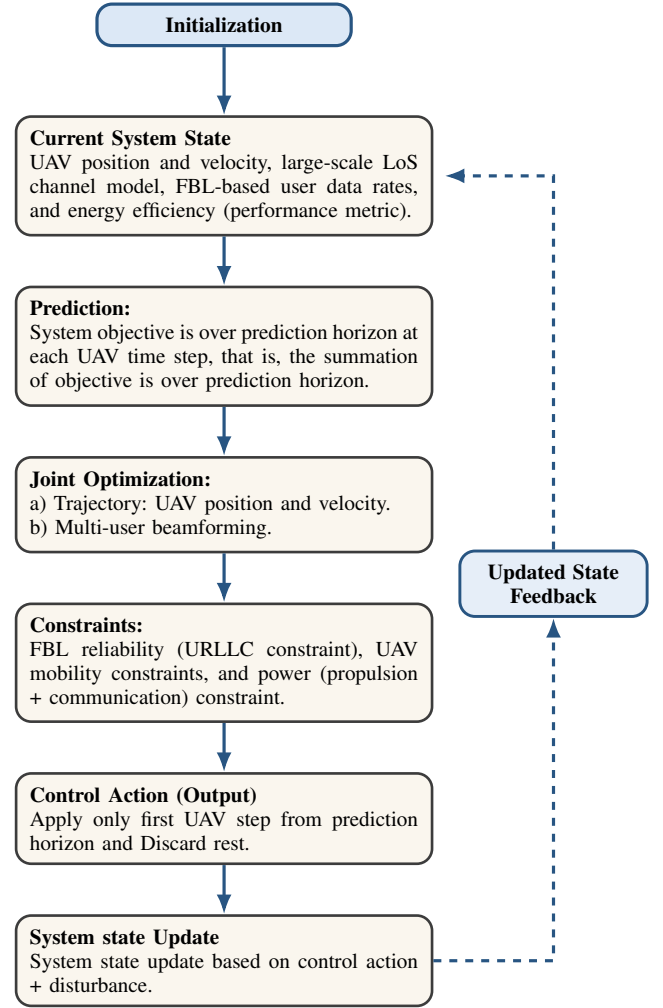

MPC enables continuous decision-making during flight by updating the UAV trajectory at each time step. As illustrated in Fig. \ref{fig:mpc_flowchart}, the UAV first uses its current state to predict future behavior over a short horizon. Then it solves an optimization problem to determine the best sequence of actions. Importantly, this optimization evaluates performance throughout the prediction horizon, thereby allowing predicted future system behavior to influence the current control decision. However, only the first action is applied. The UAV then updates its state on the basis of new information and repeats the process, forming a closed-loop operation. In this way, MPC allows the UAV to continuously adjust its trajectory, combining prediction and feedback to operate reliably in dynamic environments.
%\begin{figure}[!t]
%\centering
%\includegraphics[width=0.96\columnwidth]{Flowchart MPC.pdf}
%\caption{Flowchart of the proposed MPC-based UAV communication framework.}
%\label{fig:mpc_flowchart}
%\end{figure}
%========================================================
% Figure 2: MPC-based UAV trajectory and beamforming optimization framework
%========================================================

\section{Key Design Insights}

\subsection{Impact of FBL on System Design}
In 5G New Radio (NR), URLLC supports critical applications by transmitting short packets under strict latency and reliability constraints. Features such as mini-slot transmission reduce transmission time but also limit the available blocklength. As a result, the system operates in the FBL regime, where short-packet effects become significant \cite{11103467}. Unlike conventional Shannon-based models, where the achievable rate mainly depends on the SINR, the FBL regime introduces a reliability-related penalty term that depends on blocklength, decoding error probability, and channel dispersion \cite{11214543} (Eq. (19)). This creates a key optimization challenge because beamforming and trajectory variables affect both the achievable rate and the reliability penalty. Consequently, the optimization problem becomes highly coupled and generally non-convex, making it more difficult to solve than conventional Shannon-based formulations.

\subsection{Balancing Energy Efficiency and Mobility}
The positioning of UAVs in UAV-enabled URLLC systems affects the communication performance and the overall energy consumption. Moving closer to users improves the channel quality and reduces the transmission power, but generally cost additional maneuvering and higher propulsion energy. On the other hand, minimizing propulsion power may lead to weaker communication links and higher transmission power requirements. Meanwhile, the UAV has to keep moving towards its destination, which also limits the flexibility of its movement. Therefore, trajectory design should simultaneously balance the communication performance, system-level energy efficiency, and mission progress. The System-level energy efficiency is usually defined as the achievable data rate divided by the total UAV power consumption, including the communication and propulsion power \cite{7888557}.
\begin{figure*}[!t]
    \centering
    \includegraphics[width=0.9\textwidth]{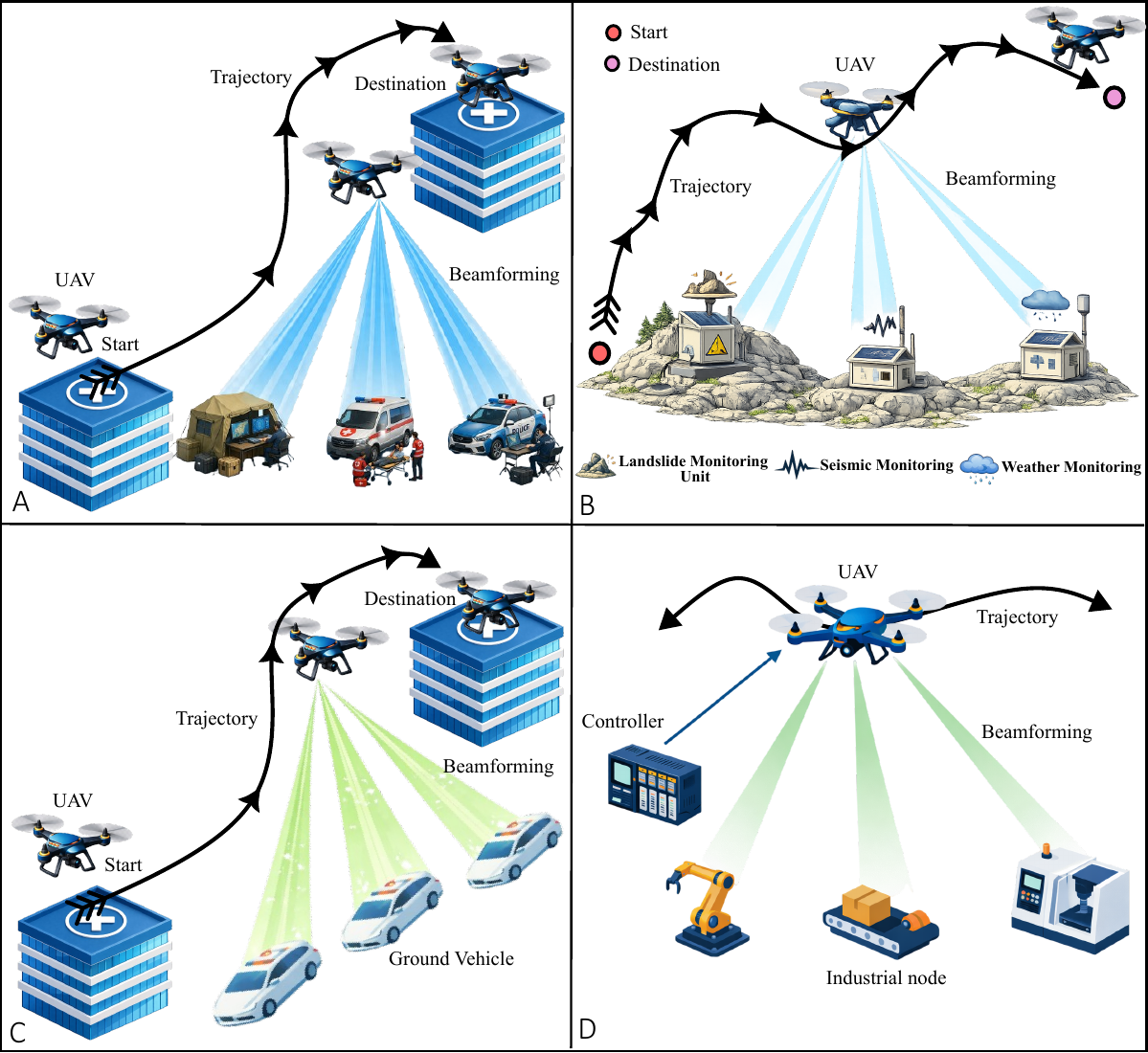}
    \caption{Application scenarios of MPC-based UAV-assisted URLLC: (A) emergency response; (B) IoT systems; (C) autonomous ground vehicles; and (D) industrial automation.}
    \label{fig:applications}
\end{figure*}
To address these coupled objectives, the system integrates energy efficiency and destination progression into a joint design objective using weighting factors \cite{9453799}. In practice, the weighting factors are selected to keep the objective terms on a comparable scale, which allows a stable balancing between energy-efficient communication and UAV trajectory progression. Then, an MPC framework is used to update the UAV trajectory and communication decisions continuously over a prediction horizon using the current system information. At each time step, only the first control action is applied before the optimization is repeated using updated system states. This receding horizon process rolls the prediction window forward at each time step with updated system information, enabling adaptive and energy-efficient UAV operation under dynamic conditions.
\section{Application Scenarios}
UAV-enabled communication can support several URLLC applications under dynamic network conditions and mobility constraints. In such scenarios, MPC enables real-time adaptation of UAV trajectory and communication decisions. Fig.~\ref{fig:applications} shows representative MPC-based UAV-assisted URLLC applications.
\subsection{MPC-based UAV-assisted URLLC for Emergency Response}
\begin{figure*}[t]
    \centering
    \includegraphics[width=0.9\textwidth]{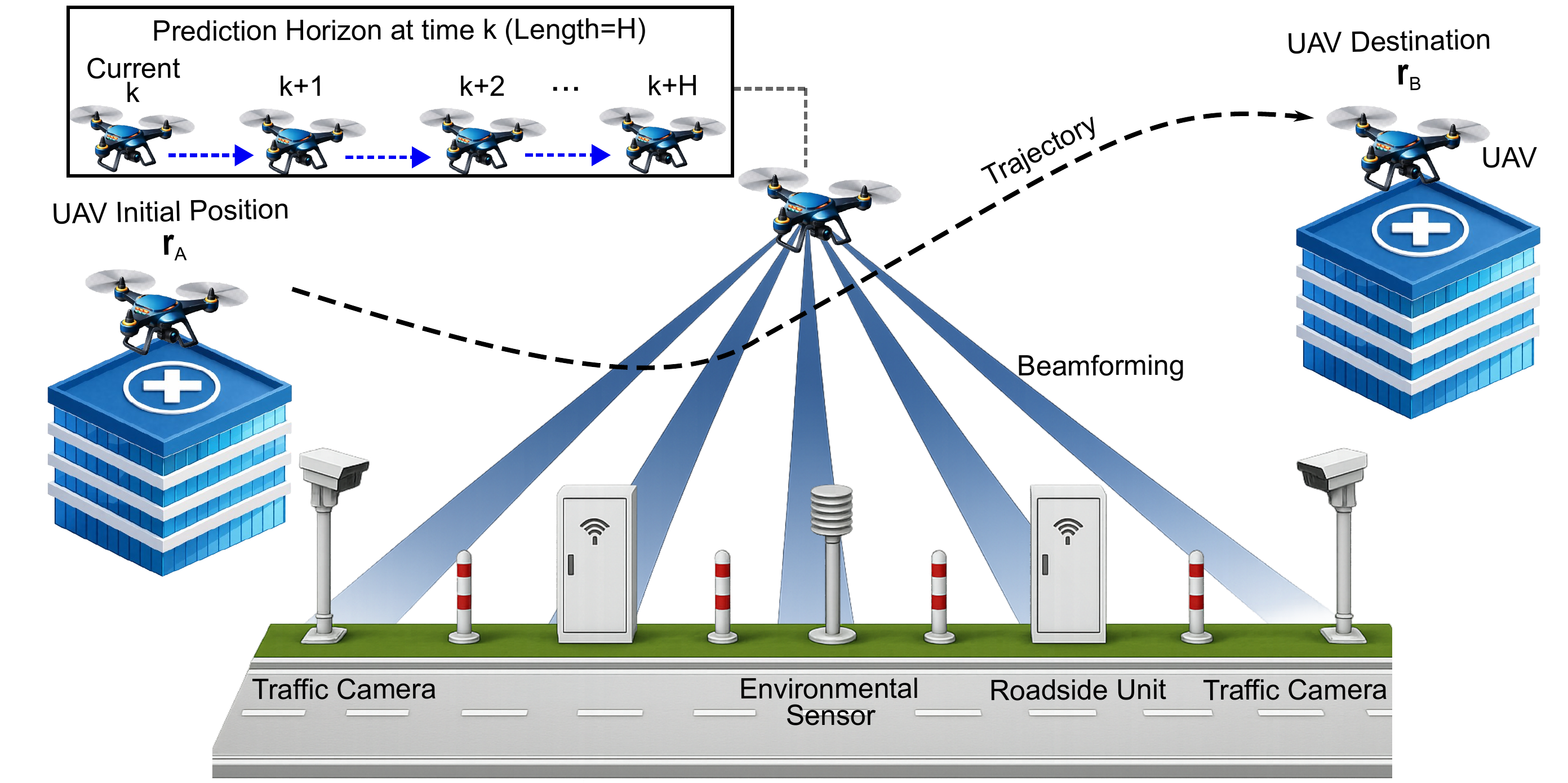}
    \caption{System model of the case study: MPC for UAV-assisted URLLC communication.}
    \label{fig:system_model}
\end{figure*}
In post-disaster situations such as earthquakes, communication infrastructure may become unavailable or severely damaged. To support rescue and coordination operations, emergency teams deploy fixed ground nodes at locations such as medical units, command centers, and security checkpoints. A UAV can provide wireless connectivity among these nodes while moving between predefined locations. The exchanged data typically consists of short mission critical packets, command directives, status updates, alarm signals, and localization information that impose high reliability and low latency. Given the spatial distribution of ground nodes, wireless channel conditions vary with distance, blockage, and propagation environment. Maintaining link quality therefore requires continuous adjustment of the UAV’s position. This mobility, however, incurs propulsion energy costs and can slow overall mission progression. An MPC-based control framework addresses this coupling by updating the UAV trajectory online using current system states and channel estimates.In this process, MPC balances link reliability, energy consumption, and UAV movement toward the target location. Recent studies on MPC-based UAV trajectory and beamforming optimization show that UAVs can adapt their positions and transmission parameters to satisfy URLLC requirements \cite{ihsan2026online}. These capabilities are important in emergency response scenarios, where communication infrastructure is damaged and reliable low-latency communication is required.

\subsection{MPC-Based UAV-assisted URLLC for IoT Systems}
In mountainous IoT monitoring systems, the distributed sensing units such as landslide, seismic and weather sensors are frequently deployed at geographically isolated locations with limited communication infrastructure. Such monitoring systems depend on reliable short-packet transmission for fault notifications, synchronization signals, and control information and therefore the system is classified as the URLLC regime under FBL transmission. Maintaining reliable connectivity from a fixed UAV position is difficult due to the severe blockage and uneven node visibility introduced by mountainous terrain \cite{9206550}. Instead, UAVs can serve as mobile aerial base stations that are dynamically repositioned depending on terrain and sensor deployment. However, frequent trajectory adjustments lead to higher energy consumption for propulsion and may be influenced by wind disturbance, localization errors and UAV motion inaccuracies. MPC tackles these challenges by adaptively optimizing UAV trajectory decisions based on the current channel and system conditions, thereby facilitating a robust and energy-efficient communication for large-scale IoT monitoring systems.

\subsection{MPC-Based UAV-assisted URLLC for Autonomous Ground Vehicles}
Autonomous vehicle systems need reliable exchange of short packets like braking commands, speed coordination signals and safety alerts under strict latency constraints. UAVs can support these systems by serving as aerial communication platforms along roadway corridors where terrestrial coverage may be insufficient. In such environments, the reliability of communication is highly dependent on the UAV positioning and the channel variations due to the mobility of the vehicles. Moreover, the trajectory tracking of UAVs and the communication performance are likely to be affected by wind disturbances, localization errors, and dynamic traffic conditions. MPC allows prediction and adaptation of the UAV trajectory and transmission parameters based on the expected vehicle motion and channel evolution over a finite horizon. Recently, distributed MPC frameworks for UAVs and vehicle platoons in multi-agent systems have been studied \cite{peng2024distributed}. Nevertheless, extending these frameworks to jointly take into account communication reliability, propulsion energy consumption and trajectory progression is an important direction for practical UAV-assisted URLLC deployment.

\subsection{MPC-Based UAV-assisted URLLC for Industrial Automation}
Industrial automation systems depend on reliable, low-latency communication between controllers and machines for motion, timing, and safety operations. In these environments, the wireless channel conditions may vary due to moving machinery and physical obstructions. The UAVs can improve the connectivity of the devices with poor link quality by adaptive positioning, as the reliability of short-packet transmission is directly determined by the UAV location \cite{ren2020joint}. However, trajectory tracking and communication reliability can be affected by localization errors, UAV motion inaccuracies, and dynamic industrial environments. The MPC continuously updates the UAV trajectory based on predicted channel conditions and real-time measurements to offer a receding-horizon framework for jointly optimizing communication performance and UAV mobility. This allows for stable short-packet communication without unnecessary movement and excess energy consumption in dynamic operation.

\section{Case Study: MPC for UAV-Assisted URLLC Communication}
\begin{figure}[!t]
    \centering

    % -------- (a) Trajectory --------
    \begin{subfigure}[t]{\columnwidth}
        \centering
        \includegraphics[width=0.85\linewidth]{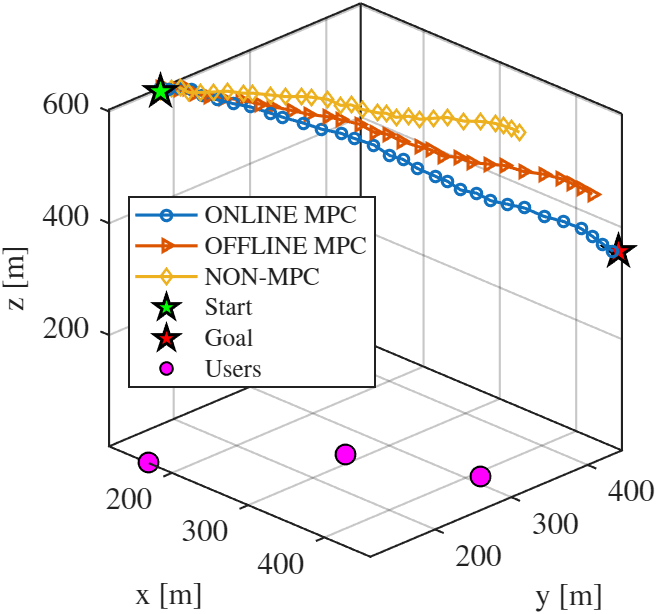}
        \caption{UAV trajectories with ground users}
    \end{subfigure}

    \vspace{0.6 em}

    % -------- (b) Power & Energy --------
    \begin{subfigure}[t]{\columnwidth}
        \centering
        \includegraphics[width=0.9\linewidth]{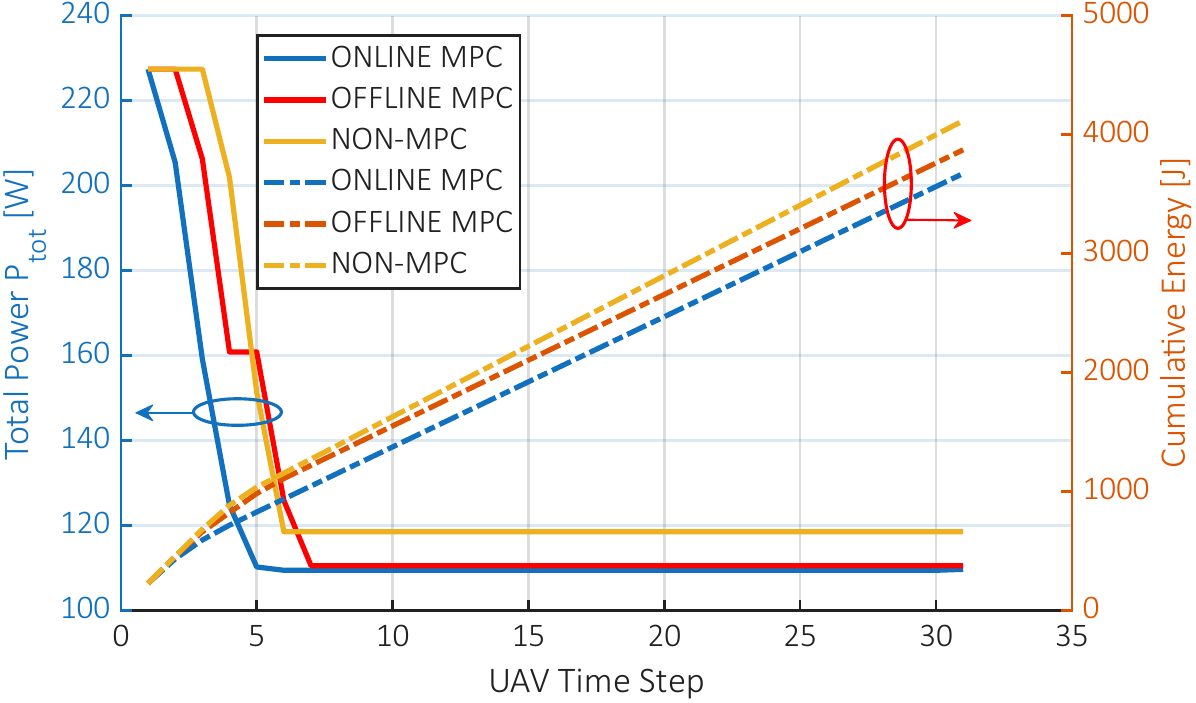}
        \caption{Total power consumption and cumulative energy}
    \end{subfigure}

    \vspace{0.6 em}

    % -------- (c) EE vs Transmission Time --------
    \begin{subfigure}[t]{\columnwidth}
        \centering
        \includegraphics[width=0.9\linewidth]{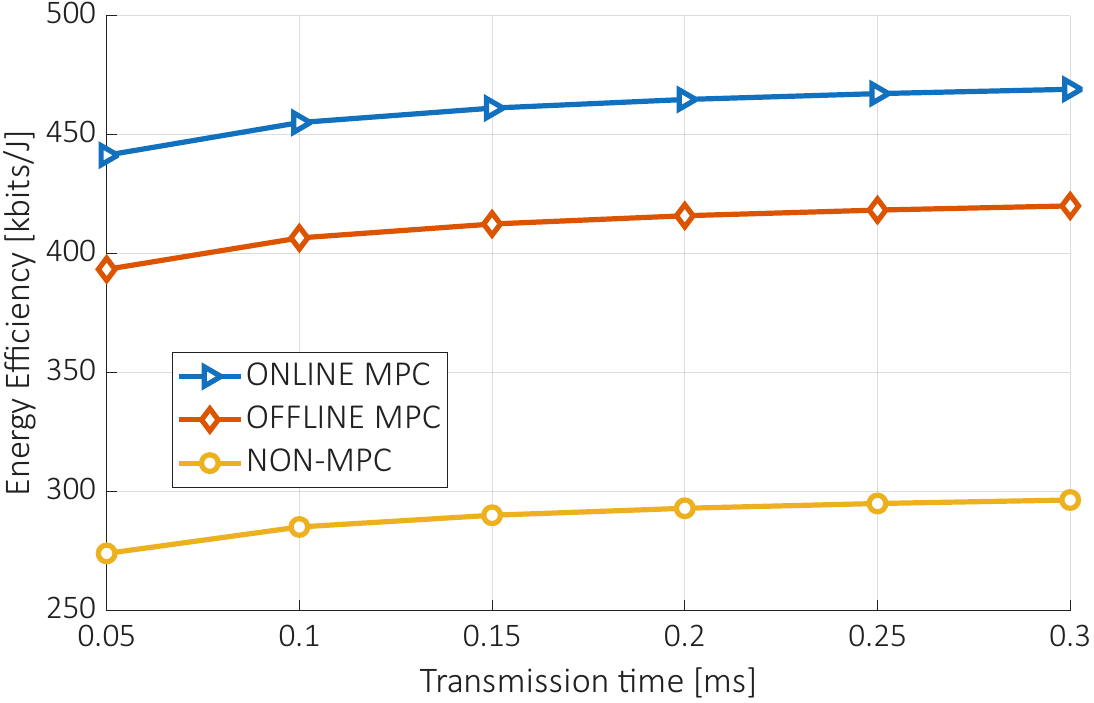}
        \caption{Energy efficiency versus transmission time}
    \end{subfigure}

    \caption{Performance comparison of online MPC, offline MPC, and non-MPC schemes.}
    \label{fig:main_results}
\end{figure}

We consider a downlink UAV-enabled communication scenario, where a UAV serving ground users moves from an initial location to a pre-defined destination, representing the corridor-based applications such as roadway monitoring, as shown in Fig. 5. The UAV provides short and delay-sensitive messages, such as control and coordination messages, with URLLC requirements. Ground users are uniformly distributed over the mission corridor. The UAV has $M=4$ antennas and communicates with $N=3$ users through a distance-dependent channel with a dominant LoS component. The system operates over a bandwidth of $5$ MHz with an FBL transmission. A baseline transmission time of $0.05$ ms is considered for joint trajectory and beamforming optimization under mobility and power constraints. The UAV motion is modeled in discrete time with a time step of 1 second and a disturbance of 5 m per step is introduced to account for practical uncertainties. We compare three approaches, online MPC, offline MPC, and a non-MPC baseline.

The resulting UAV trajectories are compared in Fig.~\ref{fig:main_results}(a). The offline MPC and non-MPC approaches compute trajectories offline and execute them without adaptation using online feedback. Thus, disturbances are accumulated over time and lead to deviation of trajectory. In contrast, the online MPC approach updates the trajectory continuously using prediction and real-time feedback, thus achieving accurate trajectory correction under disturbances. As shown in the figure, the online MPC trajectory arrives at the destination, while the offline MPC and non-MPC schemes are 86.5 m and 219.5 m away from the target, respectively.

Fig.~\ref{fig:main_results}(b) presents the total power consumption and cumulative energy usage, considering both communication and propulsion power. The online MPC approach achieves the lowest energy consumption, followed by offline MPC, whereas the non-MPC scheme exhibits the highest energy usage due to inefficient trajectory deviations under disturbances. These results show the importance of adaptive trajectory correction for energy-efficient UAV operation.

Figure~\ref{fig:main_results}(c) displays the energy efficiency versus the transmission time. In order to isolate the effect of transmission time, the trajectory of the UAV is fixed and only the beamforming variables are re-optimized. Since blocklength depends on transmission duration, shorter transmission times reduce blocklength and increase the reliability penalty under FBL communication. This reduces the achievable rate and energy efficiency. On the other hand, larger transmission times increase reliability and achievable rate, thus increasing energy efficiency. These results reflect the basic URLLC tradeoff between stringent latency requirements and communication energy efficiency.

\section{Challenges and Future Directions}
\subsection{Real-Time Complexity and Scalable Optimization}
MPC presents a predictive closed-loop control framework, but its real-time implementation in UAV-assisted URLLC systems is still computationally challenging. The complexity increases with the number of users, prediction horizon and system size which makes large scale implementation difficult in practice. Therefore, efficient implementations based on distributed and parallel processing are required in order to allow real-time operation without sacrificing control performance and reliability.

\subsection{Distributed Multi-UAV Coordination}
The generalization of the framework from a single UAV to multi-UAV scenarios raises several issues related to coordination, collision avoidance, interference avoidance, and distributed resource allocation. In such cases, the control actions become coupled not only because of the communication goals but also because of the air traffic restrictions and the dynamics of interaction between UAVs. The distributed MPC approach shows promise, but it demands robustness to asynchronicity, delays, packet loss, and limited information exchange.

\subsection{Learning-Assisted Control and Edge Offloading}
As UAV-enabled networks tend to grow increasingly complex and computationally demanding, model-based approaches alone might not be sufficient for quick adaptations. Learning-based techniques can aid the MPC process by predicting the future dynamics of channels, users demands, and the network itself, or by dynamically adjusting MPC parameters such as prediction horizon and objective function coefficients. In parallel, edge computing can help move the computational burden of optimization tasks away from the resource constrained UAVs, enabling faster and more scalable predictive control.

\subsection{Imperfect Prediction Under FBL Constraints}
A fundamental challenge in MPC-enabled UAV-assisted URLLC is that reliable control decisions depend on accurate prediction of the underlying communication and mobility states. Under FBL transmission, reliability becomes highly sensitive to mobility-induced SINR fluctuations, so errors in channel estimation, user-state prediction, blockage modeling, or localization can quickly degrade per-time-step feasibility. Future work should therefore investigate robust and stochastic MPC formulations that explicitly account for uncertainty, together with learning-assisted prediction of large-scale channel evolution and traffic demand.

\subsection{Integration with Emerging 6G Enablers}
Future UAV-assisted URLLC systems are likely to operate alongside emerging 6G enablers such as mobile edge computing reconfigurable intelligent surfaces (RIS), and movable antennas. These technologies will enhance coverage, flexibility, and reliability; however, this will be at the expense of adding another layer of complication to signaling, deployment, and coordination. A key open problem is how to co-design predictive control with such enablers while maintaining practical scalability, low latency, and energy-efficient operation.

\section{Conclusion}
This article presented the system-level overview of UAV-enabled URLLC communication, where trajectory design jointly governs both communication and propulsion energy consumption. We showed that the conventional preplanned offline approaches for joint trajectory and beamforming optimisation of UAVs often fall short in dynamic environments due to the lack of their feedback capability. In contrast, MPC provides the practical framework for real-time, predictive and feedback-driven operation of UAVs. It enables UAVs to continuously update their motion and communication decisions while balancing energy efficiency and mission progression. The discussion, case study, and application scenarios show that MPC can support robust and efficient UAV operation across a range of URLLC settings. At the same time, important challenges remain, including real-time complexity, prediction uncertainty, multi-UAV coordination, and integration with emerging 6G technologies. Addressing these issues will help move MPC-based UAV communication toward practical and scalable deployment in future wireless systems.

\bibliographystyle{IEEEtran}
\bibliography{refU}

@ARTICLE{7888557,
  author={Zeng, Yong and Zhang, Rui},
  journal={IEEE Trans. Wireless Commun.}, 
  title={Energy-Efficient {UAV} Communication With Trajectory Optimization}, 
  year={2017},
  volume={16},
  number={6},
  pages={3747--3760},
  doi={10.1109/TWC.2017.2688328}
}

@ARTICLE{9122470,
  author={Pan, Gaofeng and Lei, Hongjiang and An, Jianping and Zhang, Shuo and Alouini, Mohamed-Slim},
  journal={IEEE Trans. Wireless Commun.}, 
  title={On the Secrecy of {UAV} Systems With Linear Trajectory}, 
  year={2020},
  volume={19},
  number={10},
  pages={6277--6288},
  doi={10.1109/TWC.2020.3002230}
}

@ARTICLE{11390050,
  author={Sheemar, Chandan Kumar and Mahmood, Asad and Thomas, Christo Kurisummoottil and Alexandropoulos, George C. and Querol, Jorge and Chatzinotas, Symeon and Saad, Walid},
  journal={IEEE Trans. Commun.}, 
  title={Joint Beamforming and {3D} Location Optimization for Multi-User Holographic {UAV} Communications}, 
  year={2026},
  volume={74},
  pages={4974--4988},
  doi={10.1109/TCOMM.2026.3663285}
}

@ARTICLE{11069265,
  author={Li, Zhendong and Su, Chang and Su, Zhou and Peng, Haixia and Wang, Yuntao and Chen, Wen and Wu, Qingqing},
  journal={IEEE Trans. Commun.}, 
  title={Model Predictive Control Enabled {UAV} Trajectory Optimization and Secure Resource Allocation}, 
  year={2025},
  volume={73},
  number={11},
  pages={12652--12665},
  doi={10.1109/TCOMM.2025.3585581}
}

@inproceedings{mahmood2025uav,
  title={{UAV}-Assisted {5G} Networks: Mobility-Aware {3D} Trajectory Optimization and Resource Allocation for Dynamic Environments},
  author={Mahmood, Asad and Vu, Thang X. and Khan, Wali Ullah and Chatzinotas, Symeon and Ottersten, Bj{\"o}rn},
  booktitle={IEEE VTC2025-Fall},
  pages={1--7},
  year={2025},
  organization={IEEE}
}

@ARTICLE{11103467,
  author={Zhang, Yixin and Cheng, Wenchi and Wang, Jingqing and Zhang, Wei},
  journal={IEEE Commun. Mag.}, 
  title={Delay Trade-Off and Adaptive Resource Element Framework for {URLLC}}, 
  year={2025},
  volume={63},
  number={8},
  pages={154--160},
  doi={10.1109/MCOM.005.2400054}
}

@ARTICLE{9453799,
  author={Tuan, Hoang Duong and Nasir, Ali Arshad and Savkin, Andrey V. and Poor, H. Vincent and Dutkiewicz, Eryk},
  journal={IEEE J. Sel. Areas Commun.}, 
  title={{MPC}-Based {UAV} Navigation for Simultaneous Solar-Energy Harvesting and Two-Way Communications}, 
  year={2021},
  volume={39},
  number={11},
  pages={3459--3474},
  doi={10.1109/JSAC.2021.3088633}
}

@article{ihsan2026online,
  title={Online Model Predictive Control for Trajectory and Beamforming Optimization in {UAV}-Enabled {URLLC}},
  author={Ihsan, Asim and Asif, Muhammad and Nasir, Ali Arshad and Rabie, Khaled M. and Khan, Wali Ullah},
  journal={arXiv preprint arXiv:2603.13731},
  year={2026}
}

@ARTICLE{9206550,
  author={Ranjha, Ali and Kaddoum, Georges},
  journal={IEEE Internet Things J.}, 
  title={{URLLC} Facilitated by Mobile {UAV} Relay and {RIS}: A Joint Design of Passive Beamforming, Blocklength, and {UAV} Positioning}, 
  year={2021},
  volume={8},
  number={6},
  pages={4618--4627},
  doi={10.1109/JIOT.2020.3027149}
}

@article{peng2024distributed,
  title={Distributed Model Predictive Control for Unmanned Aerial Vehicles and Vehicle Platoon Systems: A Review},
  author={Peng, Yang and Yan, Huaicheng and Rao, Kai and Yang, Penghui and Lv, Yunkai},
  journal={Intell. Robot.},
  volume={4},
  number={3},
  pages={293--317},
  year={2024},
  publisher={OAE Publishing}
}

@article{ren2020joint,
  title={Joint Transmit Power and Placement Optimization for {URLLC}-Enabled {UAV} Relay Systems},
  author={Ren, Hong and Pan, Cunhua and Wang, Kezhi and Xu, Wei and Elkashlan, Maged and Nallanathan, Arumugam},
  journal={IEEE Trans. Veh. Technol.},
  volume={69},
  number={7},
  pages={8003--8007},
  year={2020},
  publisher={IEEE}
}

@ARTICLE{11214543,
  author={Ihsan, Asim and Asif, Muhammad and Safi, Hossein and Tavakkolnia, Iman and Haas, Harald},
  journal={IEEE Trans. Green Commun. Netw.}, 
  title={Efficient Service Differentiation and Energy Management in Hybrid {WIFI}/{LIFI} Networks}, 
  year={2026},
  volume={10},
  pages={1335--1351},
  doi={10.1109/TGCN.2025.3624594}
}

@ARTICLE{10839492,
  author={Ali, Zain and Asif, Muhammad and Khan, Wali Ullah and Elfikky, Abdelrahman and Ihsan, Asim and Ahmed, Manzoor and Ranjha, Ali and Srivastava, Gautam},
  journal={IEEE Trans. Consum. Electron.}, 
  title={Hybrid Optimization for {NOMA}-Based Transmissive-{RIS} Mounted {UAV} Networks}, 
  year={2025},
  volume={71},
  number={2},
  pages={3740--3752},
  doi={10.1109/TCE.2025.3528929}
}

@ARTICLE{9938397,
  author={Saccani, Danilo and Cecchin, Leonardo and Fagiano, Lorenzo},
  journal={IEEE Trans. Control Syst. Technol.}, 
  title={Multitrajectory Model Predictive Control for Safe {UAV} Navigation in an Unknown Environment}, 
  year={2023},
  volume={31},
  number={5},
  pages={1982--1997},
  doi={10.1109/TCST.2022.3216989}
}

@ARTICLE{9678336,
  author={Alkayas, Abdulaziz Y. and Chehadeh, Mohamad and Ayyad, Abdulla and Zweiri, Yahya},
  journal={IEEE Access}, 
  title={Systematic Online Tuning of Multirotor {UAV}s for Accurate Trajectory Tracking Under Wind Disturbances and In-Flight Dynamics Changes}, 
  year={2022},
  volume={10},
  pages={6798--6813},
  doi={10.1109/ACCESS.2022.3142388}
}

\end{document}